\documentclass[%
 reprint,
superscriptaddress,
 aps,
]{revtex4-2}

\usepackage{graphicx}
\usepackage{amssymb}
\usepackage{amsmath}
\usepackage{booktabs}
\usepackage{ulem}
\usepackage{bm}

\begin{document}

\preprint{APS/PRE}

\title{Recurrence Patterns Correlation}

\author{Gabriel Marghoti}
\email[Corresponding author: ]{gabrielmarghoti@gmail.com}
\affiliation{Physics Department, Federal University of Paraná, Curitiba, Paraná 81530-015, Brazil}

\affiliation{Potsdam Institute for Climate Impact Research, Member of the Leibniz Association, Telegraphenberg, 14473 Potsdam, Germany}

\author{Matheus Palmero Silva}

\affiliation{Institute of Mathematical and Computer Sciences, University of São Paulo, 13560-970 São Carlos, SP, Brazil}
\affiliation{Potsdam Institute for Climate Impact Research, Member of the Leibniz Association, Telegraphenberg, 14473 Potsdam, Germany}

\author{Thiago de Lima Prado}
\affiliation{Physics Department, Federal University of Paraná, Curitiba, Paraná 81530-015, Brazil}

\author{Sergio Roberto Lopes}
\affiliation{Physics Department, Federal University of Paraná, Curitiba, Paraná 81530-015, Brazil}

\author{Jürgen Kurths}
\affiliation{Potsdam Institute for Climate Impact Research, Member of the Leibniz Association, Telegraphenberg, 14473 Potsdam, Germany}
\affiliation{Humboldt Universität zu Berlin, Berlin 10099, Germany}
\affiliation{Research Institute of Intelligent Complex Systems, Fudan University, Shanghai 200433, China}

\author{Norbert Marwan}
\affiliation{Potsdam Institute for Climate Impact Research, Member of the Leibniz Association, Telegraphenberg, 14473 Potsdam, Germany}
\affiliation{Institute for Geosciences, University of Potsdam, Potsdam 14476, Germany}
\affiliation{Institute for Physics and Astronomy, University of Potsdam, Potsdam 14476, Germany}

\begin{abstract}

Recurrence plots (RPs) are powerful tools for visualizing time series dynamics; however, traditional Recurrence Quantification Analysis (RQA) often relies on global metrics, such as line counting, that can overlook system-specific, localized structures. To address this, we introduce Recurrence Pattern Correlation (RPC), a quantifier inspired by spatial statistics that bridges the gap between qualitative RP inspection and quantitative analysis. RPC is designed to measure the correlation degree of an RP to patterns of arbitrary shape and scale. By choosing patterns with specific time lags, we visualize the unstable manifolds of periodic orbits within the Logistic map bifurcation diagram, dissect the mixed phase space of the Standard map, and track the unstable periodic orbits of the Lorenz '63 system's 3-dimensional phase space. This framework reveals how long-range correlations in recurrence patterns encode the underlying properties of nonlinear dynamics and provides a more flexible tool to analyze pattern formation in recurrent dynamical systems.

\end{abstract}


\date{\today}

\maketitle

\section{\label{sec:intro}Introduction}

In time series, especially those from dynamical systems, observations close in time are typically more similar than those farther apart. This temporal dependence forms the basis of many modeling techniques and reflects a fundamental property of real-world processes: proximity in time implies relatedness. This notion closely parallels Tobler’s First Law of Geography: ``everything is related to everything else, but near things are more related than distant things”~\cite{tobler1970computer}. Although originally formulated in a spatial context, this principle applies naturally to the temporal domain, where ``nearness" refers to time lags rather than spatial distance.

A powerful framework for analyzing such temporal structures is Recurrence Quantification Analysis (RQA)~\cite{marwan2007recurrence}, which quantifies patterns in a Recurrence Plot (RP)~\cite{Eckamnn_1987_RP} -- a binary matrix representing state revisits in phase space. Traditional RQA measures, like determinism or laminarity, characterize system dynamics by counting simple structures like diagonal or vertical lines. While recent extensions have focused on quantifying more generic recurrence motifs~\cite{flauzino2025quantifying, corso2018quantifying, marghoti2024involution, hirata2021recurrence, delage2025directed}, these approaches often rely on global statistics or predefined pattern templates. This methodology faces a fundamental challenge: recurrence patterns in complex systems are not generated globally and homogeneously, but are often localized, heterogeneous outcomes of the underlying nonlinear dynamics. Consequently, existing methods struggle to capture the full diversity of patterns with arbitrary shapes and sizes, overlooking rich, localized information.

To address this gap, we draw an analogy with spatial statistics and adapt Moran’s \( I \)~\cite{moran1950notes, li2007beyond, chen2023spatial}, a classical measure of spatial autocorrelation, to the recurrence quantification analysis (RQA) framework. By using the RP as a two-dimensional grid representation of the time series, we redefine Moran's spatial weights to construct a flexible tool, the Recurrence Pattern Correlation (RPC). This quantifier is designed to detect correlations between recurrence structures in any specified direction or geometric configuration. 

Our approach differs fundamentally from standard measures such as distance correlation~\cite{szekely2007measuring} or mutual information~\cite{cover1999elements}, which operate directly on the time series values. In contrast, RPC analyzes the two-dimensional geometric relationships among recurrent states in the vicinity of each point in phase space.

The strength of this geometric approach becomes apparent when considering the organizing principles of chaotic dynamics. The complex motion on a strange attractor is governed by a hidden “skeleton” of unstable periodic orbits (UPOs) and their associated stable and unstable manifolds~\cite{gilmore2012topology, auerbach1987exploring, grebogi1988unstable}, which act as channels that guide the flow. This underlying structure leaves a distinct imprint on the RP, trajectories influenced by the same UPO tend to exhibit similar future evolution, resulting in correlated, long-range recurrence patterns that standard RQA often fails to capture. 

This paper first demonstrates that RPC behaves analogously to traditional RQA methods when standard motifs are used. It then introduces a localized, time-indexed variant of RPC as a powerful tool for probing the geometric structure of dynamical systems. This approach effectively visualizes unstable manifolds of periodic orbits in the bifurcation diagram of the Logistic map, reveals the mixed phase space of the Standard map, and highlights the UPO skeleton of the Lorenz '63 system. When motif patterns aligned with diagonal or vertical structures are employed, RPC recovers classical RQA measures, while more generic motifs provide a natural extension for capturing system-specific, localized recurrence structures.

\section{\label{sec:method}Methods}

A \textit{Recurrence Plot} (RP) is an advanced tool for nonlinear data analysis. It visualizes the times at which a dynamical system revisits a state within a specified neighborhood \(\varepsilon\).

Formally, given a time series \(\{\mathbf{x}_i\}_{i=1}^N\), the recurrence matrix  \textbf{R} is defined as:
\begin{equation}
\text{\textbf{R}}_{i,j} = \Theta(\varepsilon - \| \mathbf{x}_i - \mathbf{x}_j \|), \quad i,j = 1, \ldots, N,   
\label{eq:Rij}
\end{equation}
where \(\mathbf{x}_i\) and \(\mathbf{x}_j\) are reconstructed state space, \(\|\cdot\|\) denotes a norm (typically the Euclidean norm), and \(\Theta(\cdot)\) is the Heaviside function.

The structural patterns observed in an RP reflect the rules that govern the underlying system's dynamics. Consequently, RPs provide a robust framework for identifying complex behaviors, such as determinism, chaos, and stochasticity, in time series data through RQA.

Standard RQA quantifiers are designed to capture recurrence structures in an RP that reflect underlying dynamical behaviors. Two of the most widely used quantifiers are Determinism (DET) and Laminarity (LAM) \cite{marwan2007recurrence}. DET measures the fraction of recurrence points that form diagonal lines, indicating that nearby trajectories in phase space evolve similarly for a period of time. This is characteristic of deterministic dynamics, such as periodic or chaotic systems. On the other hand, LAM quantifies the fraction of recurrence points forming vertical lines, which correspond to laminar phases where the system’s state remains relatively unchanged over time. For example, the presence of finite-length diagonal lines in an RP is typically associated with deterministic chaos or quasi-periodic motion, reflecting sensitivity to initial conditions. In such cases, these diagonal structures are often interspersed with isolated points or other patterns that further indicate chaotic behavior.

In this work, we go beyond quantifying a small set of predefined structures, such as diagonal or vertical lines, and instead address a more fundamental question: the system’s capacity to generate structured recurrence patterns of any kind. Our goal is to develop a flexible tool that is not constrained by fixed templates but can detect arbitrary correlations emerging from the system’s underlying nonlinear dynamics

\begin{figure}
    \centering
    \includegraphics[width=\linewidth]{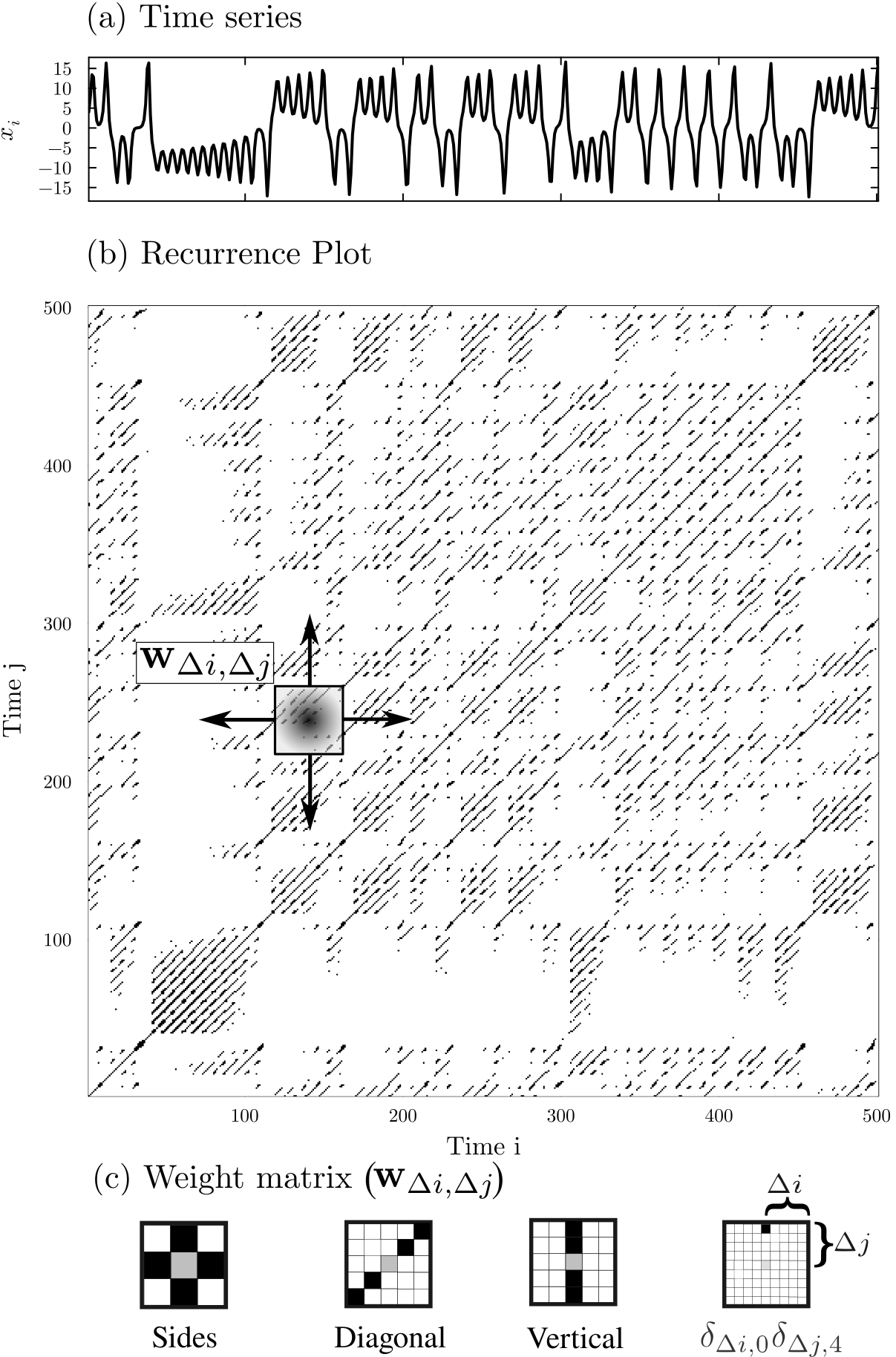}
    \caption{Recurrence Patterns Correlation method key elements. From a (a) time series, we generate a (b) standard RP, and by choosing an appropriate (c) weight matrix, we can quantify to which structure the patterns within the RP are correlated.
    Black regions indicate entries \(1\) while white regions indicate \(0\). The gray index indicates the reference recurrence point at \(\Delta i = \Delta j = 0\), which is not informative, thus set \(\text{\textbf{w}}_{0,0}=0\).}
    \label{fig:rpc_diagram}
\end{figure}

Inspired by Moran's I for measuring spatial correlation~\cite{moran1950notes, li2007beyond, marwan2007measures}, we propose a coherence measure based on recurrence structures extracted from an RP. This measure quantifies how the recurrence patterns at time \( i \) correlate with a given structure of interest, providing insight into local structural recurrence coherence from the dynamical system's time series, as illustrated in Fig. \ref{fig:rpc_diagram}. Below we provide the formal definition of the method.

We define the Recurrence Pattern Correlation (RPC) as:
\begin{equation}
\text{RPC} = 
        \sum_{\substack{i=1 \\ j=1 \\ i \ne j}}^{N} 
        \sum_{\substack{i'=1 \\ j'=1 \\ i' \ne j'}}^{N} 
        \frac{\text{\textbf{w}}_{\Delta i, \Delta j}}{W} 
        \frac{ \left(\text{\textbf{R}}_{i,j} - rr\right) \left(\text{\textbf{R}}_{i',j'} - rr\right) }{rr(1 - rr)} ,
\label{eq:rpc_definition}
\end{equation}
where we combine summations which run over the two RP indexes, \(N\) is the time series length, \( \text{\textbf{R}}_{i,j} \) is the recurrence plot element at time pair \( (i, j) \), and 
\begin{equation}
rr = \frac{1}{N^2} \sum_{i=1}^{N} \sum_{j=1}^{N} \text{\textbf{R}}_{i,j},
\label{eq:recurrence_rate}
\end{equation}
is the average over all entries, known as the recurrence rate, so the denominator is the variance of the RP matrix. The weights \( \text{\textbf{w}}_{\Delta i, \Delta j} \) are defined over time lags \( \Delta i = i' - i \) and \( \Delta j = j' - j \), so the correlation structure is translation invariant over the entire RP, making RPC a global measure of such generic motif. Moreover, the quantifier is normalized given the factor 
\begin{equation}
W = \sum_{\substack{i=1 \\ j=1 \\ i \ne j}}^{N}  \sum_{\substack{i'=1 \\ j'=1}}^{N} \text{\textbf{w}}_{\Delta i, \Delta j}.
\label{eq:W}
\end{equation}

Assuming that the relatedness of time–series values is local, we restrict the lags to a finite window so that $\mathbf{w}_{\Delta i,\Delta j}$ is non-zero only within a bounded support. This prevents boundary artifacts and keeps the
method computationally feasible even for larger weight matrices. Consequently,
the inner summation is evaluated only over indices where the weights are
non-zero. Near the boundaries of the RP, no issues arise because the
normalization factor $W$ is computed only over the admissible combinations of
time lags that fit inside the RP; outside this range, the corresponding weights
are identically zero.

Although the definition of RPC is formally equivalent to Moran's I for spatial correlation~\cite{moran1950notes}, the binary nature of RPs simplifies the expression in terms of the recurrence rate.
Specifically, the binary entries' variance simplifies the denominator to \( rr(1 - rr) \). 
All normalization factors ensure RPC values are bounded within the interval \( (-1, 1) \), just as typical correlation measures.

Moreover, the two-dimensional representation of a time series as an RP introduces directional dependencies in the association of values, which in turn affects pattern formation. This is a key distinction from the Pearson correlation coefficient~\cite{pearson1895regression}, which has a similar mathematical expression but lacks this directional structure. In the context of time series analysis, the choice of the weight matrix should reflect dynamical properties rather than spatial proximity, especially when the goal is to identify specific meaningful structural patterns.

In nonlinear dynamical systems, the evolution may appear stationary over long time scales. However, at shorter time scales—either in time or in phase space—the dynamics often exhibit strong state dependence, leading to nontrivial correlations between past and future states. As a result, recurrence patterns tend to be localized rather than globally uniform. In RPs, this manifests as the coexistence of multiple distinct structures, each reflecting the system's passage through different regions of the state space and the corresponding local dynamical behaviors.

In this context, we aim to develop a quantifier that captures not only global recurrence patterns but also those localized to individual values in the time series, and so in the state space. The local version can be conceptualized as applying the quantifier to individual columns of the RP, each associated with a specific time index \( i \) in Eq. \ref{eq:rpc_definition}.
Under a small recurrence-threshold condition, this correspondence can be approximated by a specific location in state space.

The local RPC (${\ell}\text{RPC}$) is defined as follows:

\begin{equation}
{\ell}\text{RPC}_i = 
        \sum_{\substack{j=1 \\ j \ne i}}^{N} 
        \sum_{\substack{i'=1 \\ j'=1 \\ i' \ne j'}}^{N} 
        \frac{\text{\textbf{w}}_{\Delta i, \Delta j}}{W_i} 
        \frac{ \left(\text{\textbf{R}}_{i,j} - rr_i\right) \left(\text{\textbf{R}}_{i',j'} - rr_i\right) }{rr_i(1 - rr_i)} ,
\label{eq:local_rpc_definition}
\end{equation}
where the local recurrence rate \(rr_i\) and the local normalization factor \(W_i\) are
\begin{equation}
rr_i = \frac{1}{N} \sum_{j=1}^{N} \text{\textbf{R}}_{i,j}, \  
W_i = \sum_{\substack{j=1 \\ i \ne j}}^{N}  \sum_{\substack{ i'=1 \\ j'=1}}^{N} \text{\textbf{w}}_{\Delta i, \Delta j}.
\label{eq:local_rr_w_i}
\end{equation}

In a way, the local RPC adjusts the associated mean and variance based on the density of recurrence points around each time index \(i\), thereby optimizing pattern detection under varying contrast conditions.
In all cases, we do not consider recurrences between the same time \(i=j\) or \(i'=j'\), as these are not informative.

As a result, the global RPC is not simply the average of the local values. This distinction makes the global definition better suited for identifying properties shared across the entire time series. The local RPC captures context-sensitive features by adapting to the local recurrence rate.

Furthermore, the local definition of RPC offers a crucial advantage in analyzing direction-dependent patterns. Since recurrence matrices are symmetric (\(\text{\textbf{R}}_{i,j} = \text{\textbf{R}}_{j,i}\)), any global quantifier necessarily averages the statistics of a pattern with its transpose. For example, a global count of vertical lines is identical to that of horizontal lines. In contrast, the local measure $\ell\text{RPC}$ operates on a single column i of the RP, allowing it to quantify motifs distinctly from their symmetric counterparts, i.e., from vertical motifs to horizontal ones, and thereby capturing directional information lost in global averages.

RPC and  $\ell\text{RPC}$ capture the degree to which recurrence patterns are consistent with a given structure \( \text{\textbf{w}}_{\Delta i, \Delta j} \), considering their two-dimensional configuration in the recurrence plot.
Different weight matrices offer distinct insights into the underlying local dynamics, as they associate different time lag combinations with the correlation measure.

By definition, the weight matrix can be any continuous function of time delay distances; in practice, here we use binary valued matrices, making the computational time lower as null entries do not need to be evaluated.
The binary \( \text{\textbf{w}}_{\Delta i, \Delta j} \) makes interpretation easier, facilitating the association with recurrence plots motifs \cite{corso2018quantifying, lopes2020parameter, hirata2021recurrence}.

A negative RPC indicates an anti-alignment between the recurrence structures and the weight matrix. This means that the signal exhibits structured recurrence patterns, but these patterns do not align with - and may even oppose - the structure searched by the given weight matrix.

\begin{figure}[htp!]
    \centering
    \includegraphics[width=\linewidth]{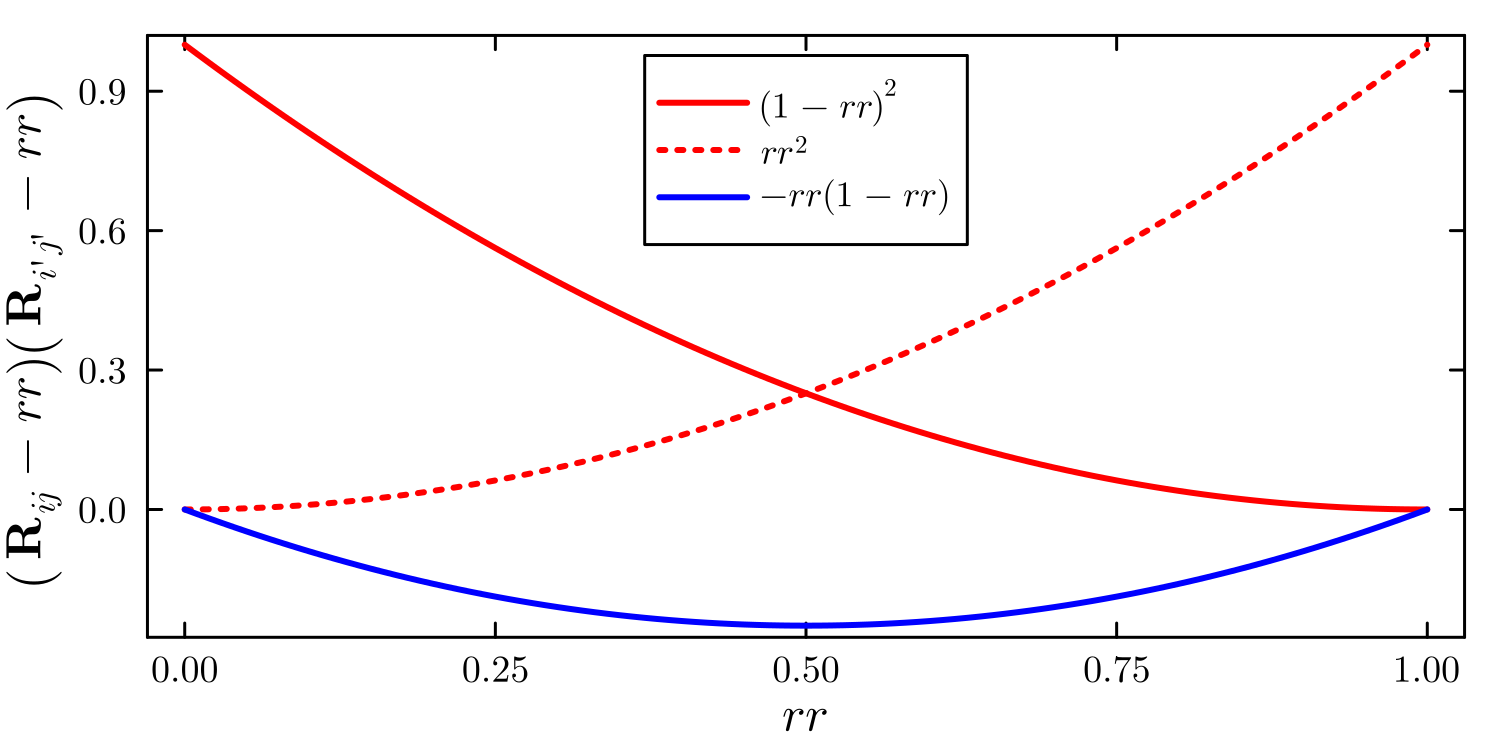}
    \caption{
        Dependence of RPC numerator terms on the recurrence rate (\(rr\)). For \(rr < 0.5\), contributions from recurring combinations, related to \((1 - rr)^2\), dominate over those from non-recurring combinations, associated with \(rr^2\). The opposite holds for \(rr > 0.5\). Consequently, the RPC quantifier becomes more sensitive to structured recurrence patterns in sparse RPs and more sensitive to non-recurrences in denser RPs. This adaptive sensitivity enhances RPC’s ability to characterize varying recurrence regimes.
        The only negative contribution to the correlation measure is the association of recurrence with no recurrence.
    }
    \label{fig:rr_functions}
\end{figure}

The numerator terms in RPC depend on the recurrence rate \(rr\) and combinations RP's binary values; therefore, such terms have only three possible outcomes: \(rr^2\), \((1 - rr)^2\), and \(-rr(1 - rr)\). There are different recurrence rate regimes in which each of such terms dominates (Fig. \ref{fig:rr_functions}). For \(rr < 0.5\), the contribution from recurring combinations, proportional to \((1 - rr)^2\), dominates over that of non-recurring combinations. Conversely, for \(rr > 0.5\), non-recurring combinations have a greater influence.

As a result, the RPC is more sensitive to structured recurrence patterns when the RP is sparse, and more sensitive to non-recurrences when the RP is dense. This adaptive behavior enhances the contrast captured by the RPC depending on the recurrence regime. 
The only combinations that negatively impact the correlation measure are those involving mismatched recurrence -- that is, the association of recurring and non-recurring time tuples. 
Moreover, the quantifier is particularly sensitive to uncorrelated structures when the RP is balanced, that is, the recurrence rate is close to \(50\%\).

The weight matrix introduces flexibility by enabling the incorporation of different underlying recurrence structures (motifs) into the correlation measure. These motifs can correspond to traditional line-based recurrence measures, such as diagonal or vertical (or horizontal) line structures, but can also capture return-time-based patterns. An example is the point-wise weight matrix \(\textbf{w}_{\delta_i,0\, \delta_j,k}\), which emphasizes recurrences for specific time lags at a fixed time series value, similar to the trajectory's return time concept~\cite{marwan2007recurrence, hirata1999statistics, penne1999dimensions}.

To make the analysis more intuitive, note that the numerator in the RPC definition can be rewritten---via a simple index swap---as a linear combination of the $\mathbf{w}_{\Delta i, \Delta j}$ terms. Consequently, the correlation for any specific recurrence pattern corresponds to the average correlation of the non-zero $\mathbf{w}_{\Delta i, \Delta j}$ terms, which are associated with generalized return times.

Therefore, most of the results we present are for cases where the weight matrix \( \text{\textbf{w}}_{\Delta i, \Delta j} \) reduces to a single nonzero entry, i.e., \( \text{\textbf{w}}_{\Delta i, \Delta j} = \delta_{\Delta i , \Delta i_0} \delta_{\Delta j, \Delta j_0} \). This corresponds to selecting a single component of the full expression, enabling a more tractable investigation of the quantifier's properties.

\section{\label{sec:results}Results and Discussion}

To analyze how the underlying dynamics impact the generation of structured patterns in an RP, we examine a set of representative dynamical systems that encompass a range of stochastic, periodic, and chaotic behaviors. The systems include:

\begin{itemize}
    \item \textbf{Gaussian white noise (GWN)}: A stochastic process with zero mean and unit variance, independently sampled at each time step from a Gaussian distribution. GWN serves as the benchmark for uncorrelated data.

    \item \textbf{Autoregressive models [AR(1)]}: Linear stochastic processes defined by
    \[
        x_{n+1} = \alpha x_n + \eta_n,
    \]
    where \(\alpha\) is the auto-regressive coefficient and \(\eta_n\) is Gaussian white noise with zero mean and unit variance. We consider two cases: \(\alpha = 0.8\) (moderate autocorrelation) and \(\alpha = 0.99\) (strong autocorrelation).

    \item \textbf{Logistic map \cite{may1976simple}}: A discrete-time nonlinear system given by
    \[
        x_{n+1} = r x_n (1 - x_n),
    \]
    generating chaotic dynamics for control parameter \(r = 4.0\).

    \item \textbf{Standard map (Chirikov–Taylor map) \cite{chirikov1979universal}}: A 2D area-preserving chaotic map defined by
    \[
    \begin{aligned}
        y_{n+1} &= x_n + K \sin(x_n) \mod 2\pi, \\
        x_{n+1} &= x_n + y_{n+1} \mod 2\pi,
    \end{aligned}
    \]
    where \(K =  2.5 \) is the nonlinearity parameter, which generates mixed phase space with chaotic or periodic orbits depending on initial conditions. 
    
    \item \textbf{Lorenz'63 system \cite{lorenz1963deterministic}}: A continuous-time chaotic system governed by
    \[
    \begin{aligned}
        \dot{x} &= \sigma(y - x), \\
        \dot{y} &= x(\rho - z) - y, \\
        \dot{z} &= x y - \beta z,
    \end{aligned}
    \]
    with standard parameters \(\sigma = 10\), \(\rho = 28\), and \(\beta = 8/3\). The system was integrated using a Runge–Kutta 4th-order method with time step \(dt = 0.01\).

    \item \textbf{Periodic sine function}: A deterministic periodic signal, \( x(t) = \sin(2\pi t)\) sampled at steps \(dt = 0.01\), with a period every \(100\) data points.
\end{itemize}

For one-dimensional time series, we applied the \textit{PECUZAL} algorithm \cite{PECUZAL_EMB_kramer2021unified} for phase space reconstruction. This optimizes embedding parameters using non-uniform delays and a cost function tailored to the embedding dimension, without relying on a threshold parameter.

\begin{figure}[htbp]
    \centering
    \includegraphics[width=\linewidth]{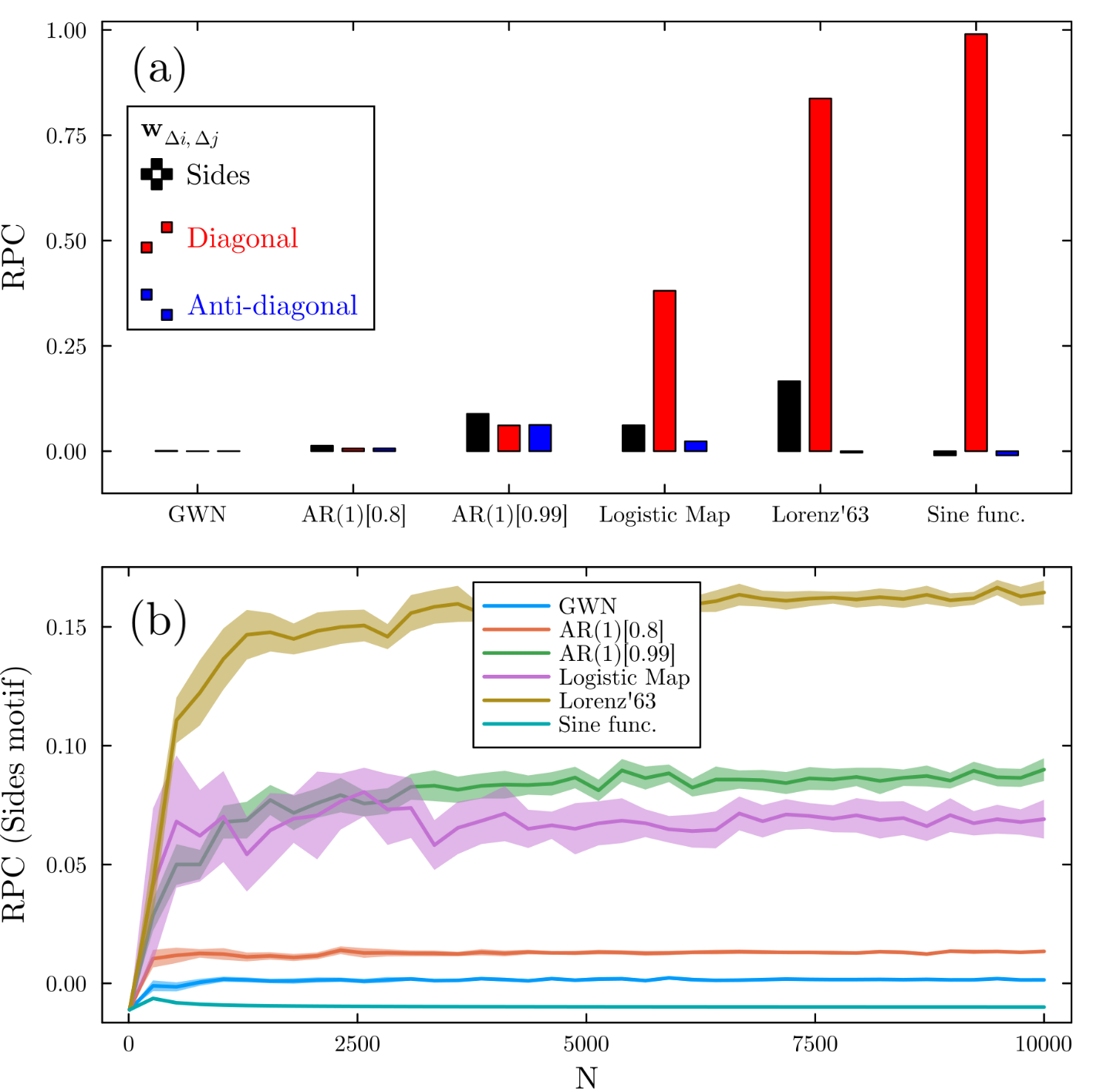}
    \caption{Recurrence Patterns Correlation (RPC) for paradigmatic dynamics and (a) representative recurrence motifs, and (b) as a function of the time series length. Here we consider a \(1\%\) recurrence rate across all recurrence plots. In this scale, the recurring condition is highlighted over nonrecurring patterns.
    Each bar represents a specific weight matrix \( \text{\textbf{w}}_{\Delta i, \Delta j}\), corresponding to: sides (\( \text{\textbf{w}}_{\Delta i, \Delta j} = \delta_{\Delta  i,0} \delta_{\Delta  j,\pm 1} + \delta_{\Delta  i, \pm 1} \delta_{\Delta  j, 0}\)), diagonals (\( \text{\textbf{w}}_{\Delta i, \Delta j} = \delta_{\Delta  i,\pm 1} \delta_{\Delta  j,\pm 1} \)), and anti-diagonals (\( \text{\textbf{w}}_{\Delta i, \Delta j} = \delta_{\Delta  i,\pm 1} \delta_{\Delta  j,\mp 1} \)).
    In the bottom panel, the solid line represents the mean over 20 trials, and the error bars indicate the corresponding standard deviation.
    In the top panel, we present results for time series with length \(N=10000\).}
    
    \label{fig:bar_plot_RPC}
\end{figure}

The simplest correlated patterns arise from combinations of nearby recurrences, defined by the weight matrix \(\text{\textbf{w}}_{\Delta i, \Delta j} = 1\) only when \(\Delta i = \pm 1\) or \(\Delta j = \pm 1\). This corresponds to single time-lag recurrences aligned with vertical, diagonal, or anti-diagonal directions. Such a definition is commonly used in spatial correlation analysis, where long-range temporal causality is disregarded in favor of identifying nearby values in spatial clusters. We apply this simple weight matrix to representative dynamical systems [Fig.~\ref{fig:bar_plot_RPC}(a)], revealing distinct recurrence fingerprints that differentiate stochastic from deterministic dynamics.

We investigated how the RPC quantifier depends on the amount of available data by computing it for increasing time series lengths and averaging over 20 independent trials [Fig. \ref{fig:bar_plot_RPC}(b)]. For sufficiently long time series, the RPC converges to a stable value with low standard deviation, whereas short series exhibit significant variability due to insufficient sampling. This analysis demonstrates that the method requires a minimal data length to obtain statistically reliable estimates and clarifies the conditions under which the measure approaches its asymptotic value.

\textbf{Diagonal Motif.} The correlation for the diagonal motif ($\text{\textbf{w}}_{\Delta i, \Delta j} = \delta_{\Delta i,\pm 1}\delta_{\Delta j,\pm 1}$) serves as a measure analogous to the RQA quantifier DET. It quantifies the prevalence of diagonal lines in an RP, which are characteristic of deterministic dynamics. Consequently, its value is negligible for Gaussian White Noise (GWN) but increases substantially for systems with memory, peaking for the deterministic chaotic and periodic systems (Logistic Map, Lorenz'63, and Sine function).

\textbf{Sides Motif.} The ``sides'' motif ($\text{\textbf{w}}_{\Delta i, \Delta j} = \delta_{\Delta i,0}\delta_{\Delta j,\pm 1} + \delta_{\Delta i,\pm 1}\delta_{\Delta j,0}$) probes for vertical and horizontal lines, which correspond to laminar states or persistence in a region of phase space. This correlation is most pronounced for the strongly correlated AR(1) process and the chaotic systems. This indicates that in these systems, states have a tendency to recur in the same neighborhood for short periods, a direct consequence of a short-term linear memory.

\textbf{Anti-Diagonal Motif.} The anti-diagonal motif ($\text{\textbf{w}}_{\Delta i, \Delta j} = \delta_{\Delta i,\pm 1}\delta_{\Delta j,\mp 1}$) provides the clearest distinction between system types. Such a pattern implies a form of time-reversal symmetry in recurrences ($x_i \approx x_{j+1}$ and $x_{i+1} \approx x_j$), a structure that is dynamically forbidden in autonomous, forward-evolving deterministic systems. Accordingly, the RPC value for this motif is nearly zero or slightly negative for all deterministic systems analyzed. In contrast, the stochastic AR(1) process exhibits a small but positive correlation. This does not imply a specific dynamical rule but rather that random fluctuations can coincidentally produce such patterns, whereas deterministic rules actively suppress them.

\begin{figure*}[!htb]
    \centering
    \includegraphics[width=0.95\linewidth]{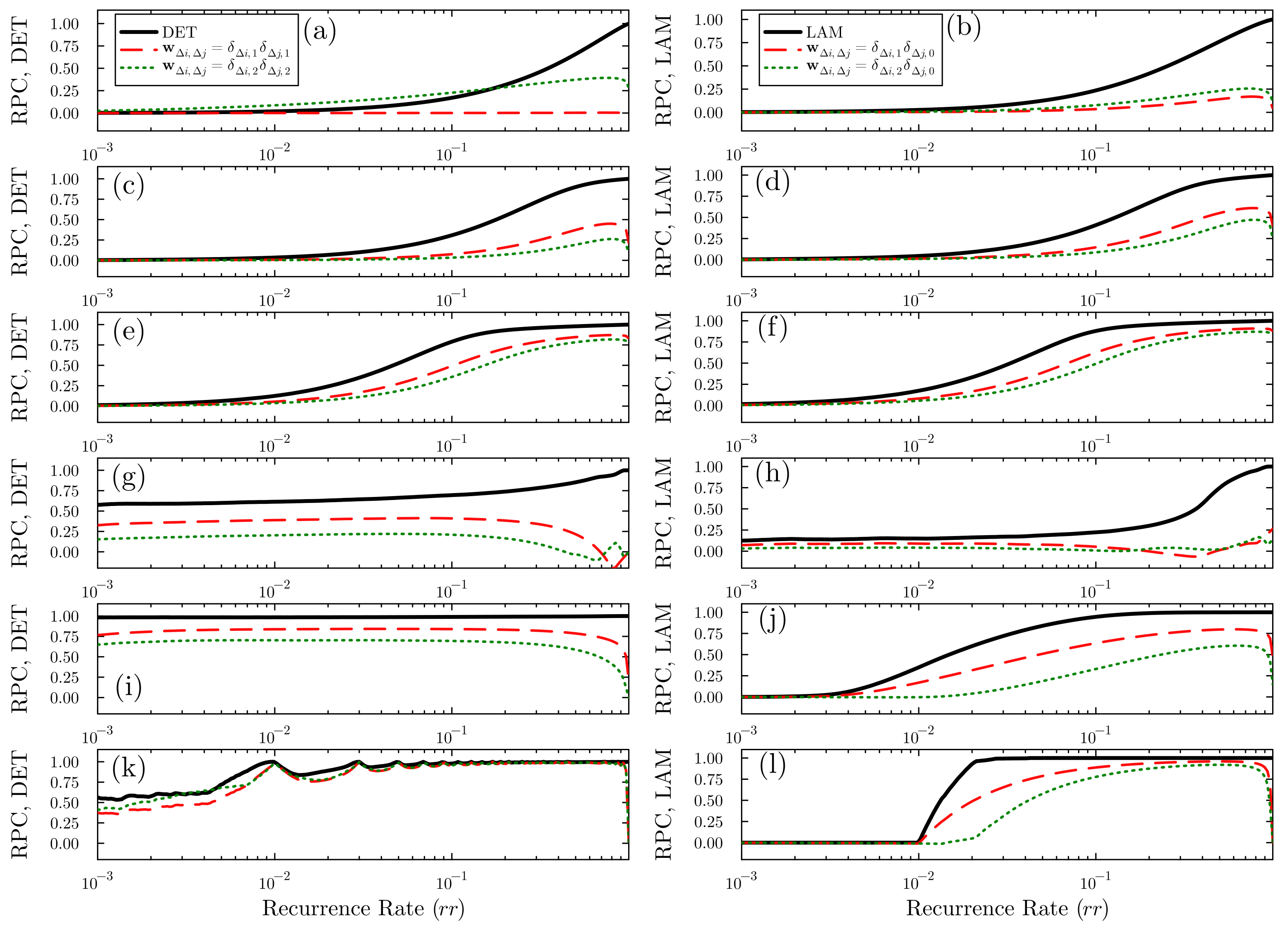}
    \caption{RPC for single structured time lags and representative dynamical systems as a function of the recurrence rate, we compare point-wise motifs on diagonal direction to DET (left) and point-wise motifs on horizontal direction to LAM (right).
    The analyzed systems are: (a-b) Gaussian white noise, (c-d) auto-regressive model with \(\alpha = 0.8\), (e-f) auto-regressive model with \(\alpha = 0.99\), (g-h) chaotic Logistic map, (i-j) chaotic Lorenz '63 system, and (k-l) periodic sine function.}
    \label{fig:RPC_vs_rr}
\end{figure*}

In summary, while both deterministic and stochastic processes with memory can generate correlated recurrence patterns, their nature is fundamentally different. Deterministic systems exhibit strong preferences, generating specific geometric structures while forbidding others. Correlated noise, however, produces correlations that are not preferentially aligned with any single geometric motif, reflecting the indiscriminate nature of its linear memory for recurrence pattern generation.

The RPC depends not only on the motif pattern structure but also on the recurrence threshold. Therefore, we analyze its performance as a function of the RP recurrence rate (Fig.~\ref{fig:RPC_vs_rr}) for representative dynamical systems. This analysis employs delta weight matrices \(\mathbf{w}_{\Delta i, \Delta j}\) with non-zero entries along the diagonal or vertical directions, evaluated at two distinct time shifts. Such weight matrices are associated with traditional RQA quantifiers, namely DET and LAM.

For GWN [Fig. \ref{fig:RPC_vs_rr}(a-b)], our quantifier detects a small amount of correlation for high recurrence rates. We attribute this finding to the non-uniformity of the Gaussian distribution itself, an issue also present on DET and LAM quantifiers. However, RPC still get a smaller correlation measure for all recurrence rates in comparison with DET and LAM, which are not suitable for high threshold values, and so high recurrence rates.

In the case of correlated noise from an AR(1) process [Fig. \ref{fig:RPC_vs_rr}(c-f)], all patterns yield similar correlation values despite the motif used. However, the correlation decreases at longer time shifts, which is consistent with the process's linear memory decay. This suggests that stochastic time series, even those with correlation, do not generate a preferred recurrence pattern.

In contrast, deterministic dynamical systems exhibit a more complex RPC profile, characterized by higher correlation for diagonal structures, particularly at low recurrence rates. For chaotic systems, this correlation decays as the time shift increases. Conversely, for periodic systems, the correlation remains maximal regardless of the time shift or recurrence rate.

A key advantage of RPC over traditional RQA is its handling of non-recurrences. For high recurrence rates ($rr > 0.5$), the contribution of non-recurring point pairs (\(\text{\textbf{R}}_{i,j}=0,\ R_{i',j'}=0\)) dominates the correlation calculation (Fig. \ref{fig:rr_functions}). This means that structured patterns of non-recurrence become informative, making RPC uniquely suitable for analyzing dense recurrence plots where traditional quantifiers often fail, as shown in Fig. \ref{fig:RPC_vs_rr} for \(rr \lessapprox 1\).

A cornerstone of chaos theory is the idea that a chaotic attractor is structured by an infinite set of Unstable Periodic Orbits (UPOs). These UPOs form the ``skeleton" of the dynamics~\cite{gilmore2012topology, auerbach1987exploring, grebogi1988unstable}. We hypothesize that the long-range temporal correlations captured by RPC are directly related to these UPOs. Moreover,  $\ell\text{RPC}$ can detect proximity to such orbits through their effect on recurrence patterns.

\begin{figure}[htbp]
    \centering
\includegraphics[width=\linewidth]{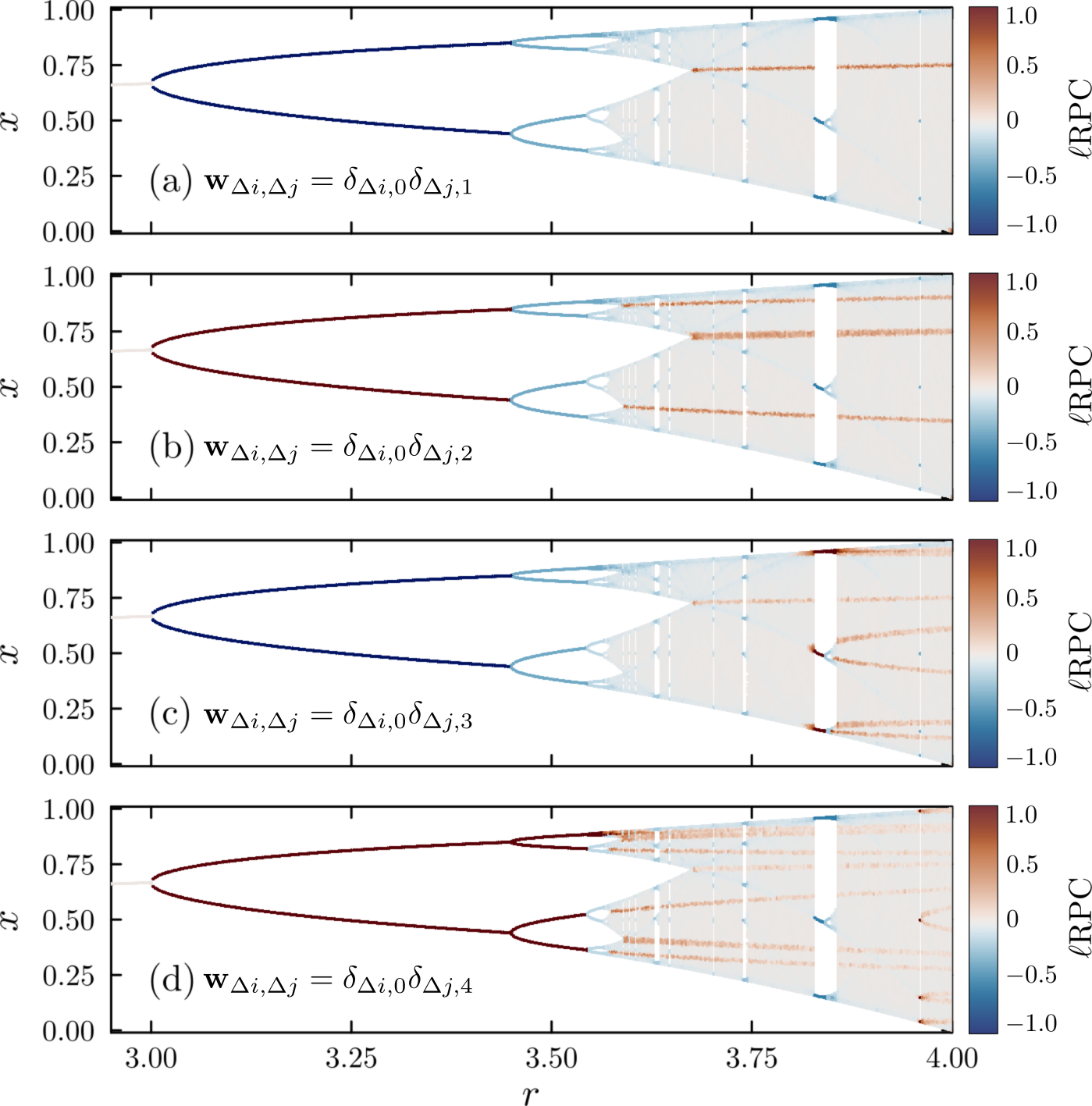}
    \caption{Logistic map bifurcation diagram along with  $\ell\text{RPC}$. We consider the simple single recurrence pattern with lag \(k\) along a vertical direction, \( \text{\textbf{w}}_{\Delta i, \Delta j} = \delta_{\Delta i, 0} \delta_{\Delta j, k}\), emphasizing the return time to a phase space position.
    We detect UPO nearby to the trajectory, related to (a) period 1, (b) period 2, (c) period 3, and (d) period 4.}
    \label{fig:logistic_map_bifurcation}
\end{figure}

The Logistic Map provides a canonical example of the transition from simple periodic behavior to chaos. We leverage the local  $\ell\text{RPC}$ to visualize and quantify the structures within its bifurcation diagram. To do this, we compute the  $\ell\text{RPC}$ for trajectories generated across a range of the control parameter $r \in [2.9, 4.0]$, from a period \(1\) stable dynamics to chaotic behavior. The results are presented in Fig. \ref{fig:logistic_map_bifurcation}, where the attractor for each $r$ value is colored according to its local  $\ell\text{RPC}$ value (Eq. \ref{eq:local_rpc_definition}).

The analysis employs a vertical line motif, specified by the weight matrix $\text{\textbf{w}}_{\Delta i, \Delta j} = \delta_{\Delta i, 0} \delta_{\Delta j, k}$, which tests for correlations at a specific time lag $k$. This effectively probes for periodicity of order $k$: it measures how strongly a state's recurrence at time $j$ is correlated with a recurrence in the same state space neighborhood at time $j+k$.

The $\ell\text{RPC}$, applied to the Logistic Map's bifurcation diagram (Fig.~\ref{fig:logistic_map_bifurcation}), effectively functions as a ``dynamical filter,'' selectively revealing the influence of underlying UPOs on the system's recurrence structure. By setting the motif's time lag to match the period of a specific UPO, the  $\ell\text{RPC}$ highlights the regions of phase space governed by that orbit. For instance, a lag of $k=2$ [Fig.~\ref{fig:logistic_map_bifurcation}(b)] yields high positive correlation values precisely along the branches of the period-2 orbit. Similarly, lags of $k=3$ and $k=4$ successfully identify the narrow period-3 window and the period-4 orbits, respectively, even deep within the chaotic regime.

Complementing this, the analysis also reveals significant \textit{negative} correlations. When the tested motif's periodicity does not match the local dynamics of a trajectory, the  $\ell\text{RPC}$ becomes negative. This indicates a dynamical avoidance of that specific recurrence pattern, a hallmark of deterministic rules. For example, in the period-2 window, testing for period-3 correlations yields strongly negative values.

We note that because our method relies on RPs, detecting a specific type of dynamics requires that (i) the system changes its states during its evolution - so that the RP contains informative structures - and (ii) the trajectory visits the phase-space region where the corresponding property is present. This explains why the Logistic Map exhibits a low $\ell\text{RPC}$ in the period-one regime ($r < 3.0$) and, similarly, why the RPC does not detect a period-one UPO when the system is currently in a higher-period periodic orbit. 

Taken together, these results demonstrate that the $\ell\text{RPC}$ can not only detect the presence and influence of UPOs but also quantify the degree to which certain recurrence patterns are either preferred or suppressed by the deterministic evolution.

\begin{figure*}
    \centering
    \includegraphics[width=0.9\linewidth]{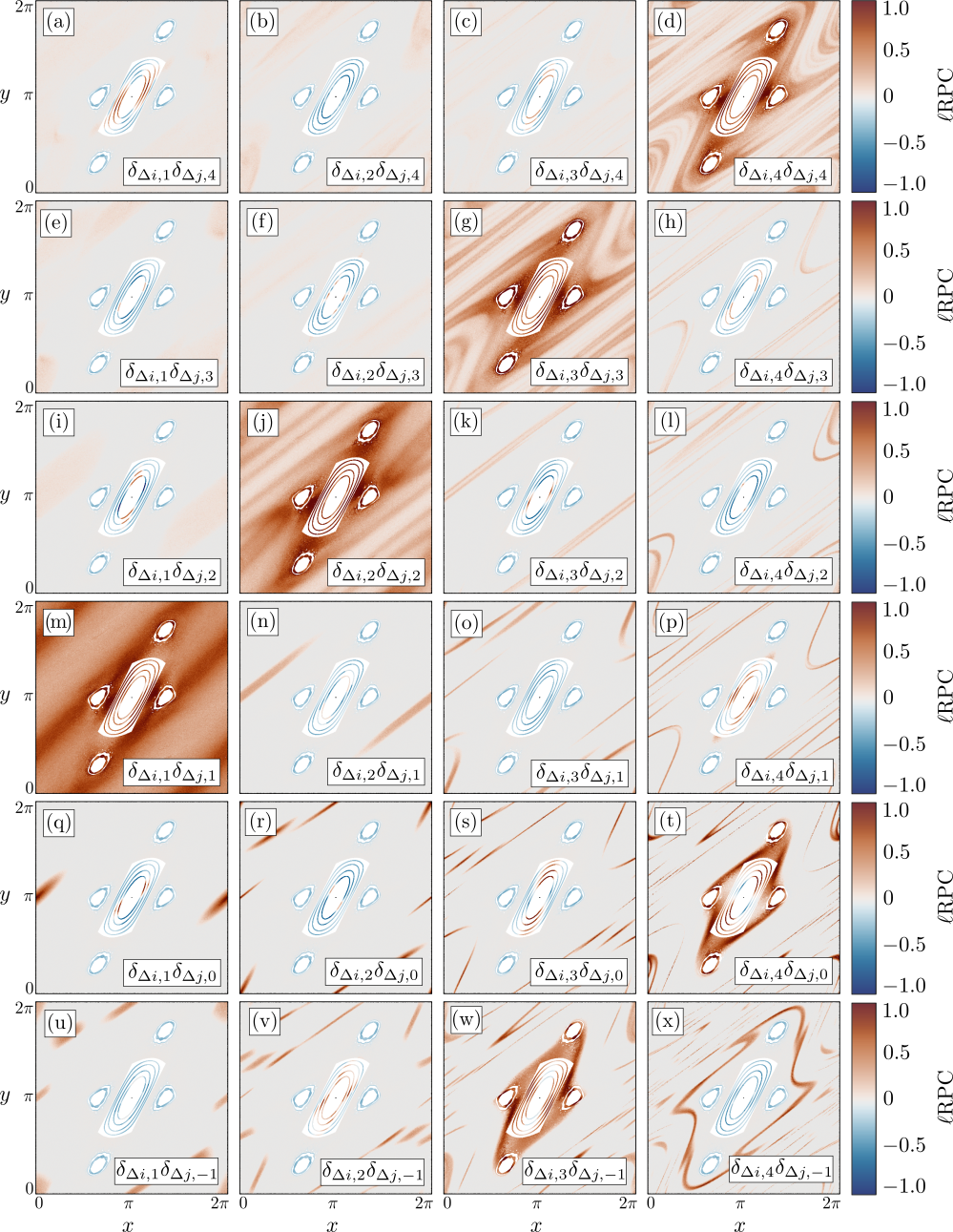}
    \caption{Local Recurrence Pattern Correlation ($\ell\text{RPC}$) in the Standard Map phase space, considering single time-lag motifs. The  $\ell\text{RPC}$ is calculated for uniformly distributed initial conditions $(x_0,y_0) \in [0,2\pi]$ across the phase space for $K=2.5$ and, with RPs built using a fixed \(\varepsilon=0.5\). Each panel shows the  $\ell\text{RPC}$ for a different correlation motif defined by the weight matrix $\text{\textbf{w}}_{\Delta i, \Delta j} = \delta_{\Delta i, k} \delta_{\Delta j, l}$, where the time lags $(k,l)$ are indicated in the panel titles. Warm colors (red) denote high positive correlation, indicating that the chosen recurrence pattern is prevalent, while cool colors (blue) denote low or negative correlation -- the chosen pattern is rare.}
    \label{fig:standard_map_panel}
\end{figure*}

To explore how $\ell\text{RPC}$ can characterize systems with a mixed phase space, we apply it to the Standard Map. The map's phase space is composed of a chaotic sea interspersed with islands of stability (periodic and quasi-periodic orbits). We calculate the $\ell\text{RPC}$ for various initial conditions $(x, y)$ spanning most of the possible orbits within the modulated phase space. The results, shown in Fig.~\ref{fig:standard_map_panel}, reveal that the choice of the weight matrix $\text{\textbf{w}}_{\Delta i,\Delta j}$ acts as a selective filter, highlighting different geometric and dynamical features, even for initial conditions with differing qualitative dynamics.

The  $\ell\text{RPC}$ method effectively dissects the mixed phase space of the Standard Map by acting as a dynamical filter (Fig.~\ref{fig:standard_map_panel}). We focus on a weight matrix with point-wise non-zero entries that takes the form $\text{\textbf{w}}_{\Delta i,\Delta j} = \delta_{\Delta i,k} \delta_{\Delta j,l}$ for various lags $k$ and $l$ combinations. This pattern probes the correlation between a recurrence at $(i, j)$ and another recurrence at $(i+k, j+l)$.

We display each panel in Fig.~\ref{fig:standard_map_panel} so that its position matches the time–lag tuple $(\Delta i, \Delta j)$ it represents in the recurrence plot. In this arrangement, motifs associated with diagonal patterns (i.e., equal time lags) are aligned along the main diagonal of the figure, while motifs associated with vertical patterns are aligned vertically. This layout provides an intuitive visual mapping between a subplot’s position and the recurrence structure it encodes.

High positive $\ell\text{RPC}$ values (in red) are predominantly found within the chaotic sea and the periodic islands, most notably for the diagonal condition \(l=k\) parallel to the main. This indicates that in chaotic regions, the deterministic evolution correlates recurrences at diagonal structures almost uniformly in the state space. If a state $x_i$ is close to $x_j$, it is highly probable that their future iterates $x_{i+k}$ and $x_{j+k}$ will also be close, \textit{independently} of their position in the state space. 
In contrast, for other combinations of time lag, the system is state-sensitive for the generation of recurrence patterns. 

The boundaries of the stability islands and the invariant KAM tori are traced by contours of high correlation. This demonstrates that $\ell\text{RPC}$ can identify regions of phase space where trajectories, though separated in time, consistently revisit the same local areas, thus mapping out the persistent geometric features of the dynamics.
In particular, the highlighted structures in phase space seem to match unstable manifolds \cite{Meiss1992, Meiss2008}, indicating that such manifolds are generators of specific recurrence structures in the data at time lags equal to their associated UPO period.

All possible combinations of time lags exhibit correlated patterns associated with specific regions of phase space. These patterns are particularly pronounced when the lags share the same total number of iterations, \(l+k\).
This behavior arises from the deterministic nature of the system, which induces recurrence structures along diagonals. However, unlike traditional RQA quantifiers, these patterns emerge along diagonals that are parallel to the reference recurrence time tuple, rather than directly aligned with it.
Under such conditions, the generation of recurrence patterns becomes more sensitive to the system’s state than the standard diagonal-aligned recurrences [Fig. \ref{fig:standard_map_panel}(d,g,j,m)].

The regular islands of stability are delineated by  $\ell\text{RPC}$ patterns. For motifs with time lags that do not match the island's intrinsic periodicity, these regions exhibit low or even negative correlation. This occurs because their rigid, quasi-periodic dynamics are incompatible with the tested recurrence pattern. Conversely, motifs precisely tuned to an island's period can generate high positive correlations, selectively highlighting its structure [i.e., Fig. \ref{fig:standard_map_panel}(t)].

\begin{figure}[htbp]
    \centering
    \includegraphics[width=\linewidth]{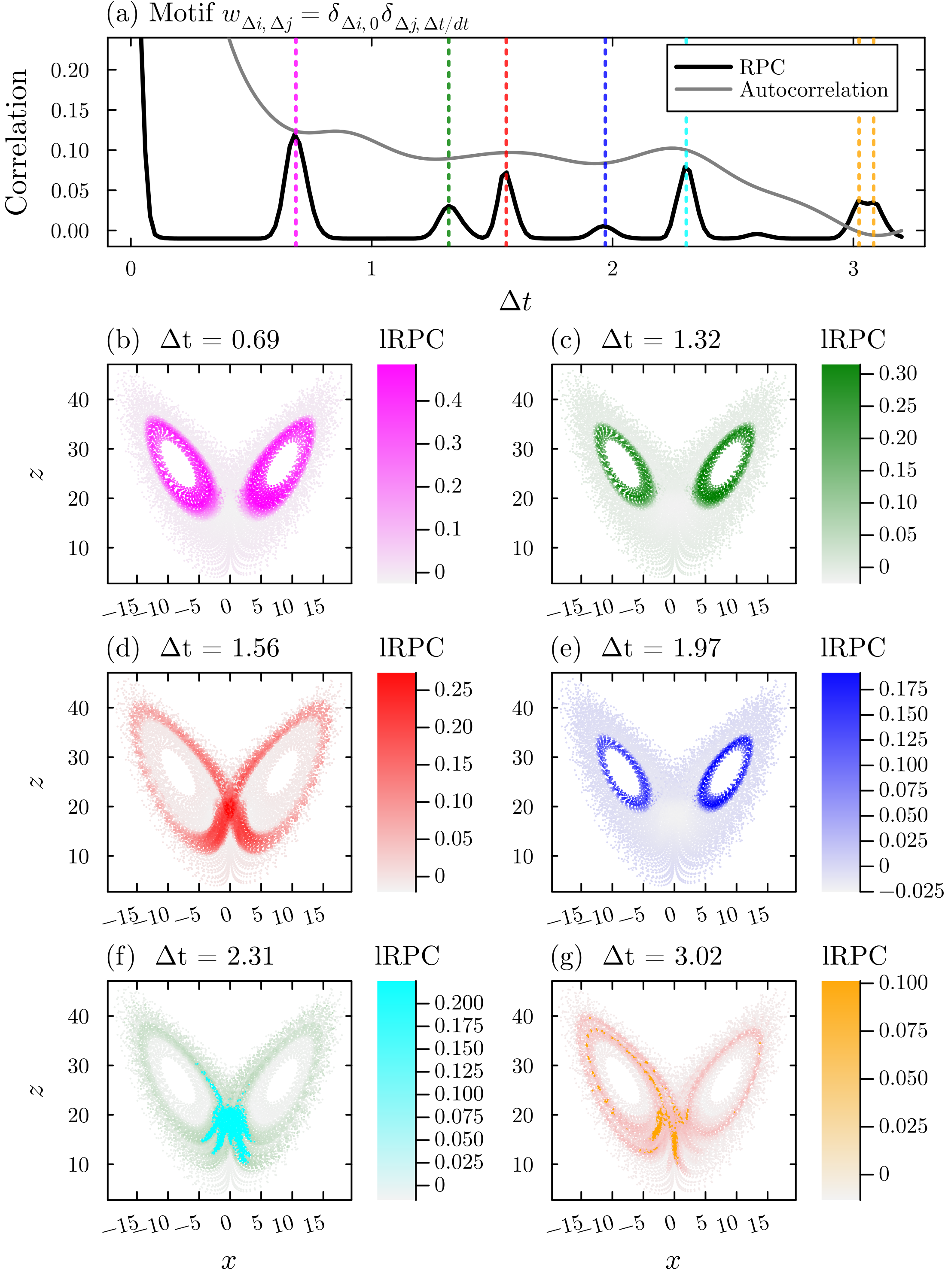}
\caption{Detection of inherent recurrence times using $\ell\text{RPC}$ in the Lorenz '63 system with a sampling step of \(dt = 0.02\).  
Panel (a) shows the global RPC as a function of time lag \(\Delta t\), alongside the component \(x\) autocorrelation function for comparison.  
The bottom panels depict the localized $\ell\text{RPC}$ over the Lorenz attractor, evaluated at time lags corresponding to prominent peaks in the RPC. Some peak lags coincide with the durations of known heteroclinic unstable periodic orbits (UPOs), indicated by vertical lines colored by period~\cite{LI2024114751_UPO_Lorenz}.
We highlight three UPOs with the shortest durations: (d) period-2 (LR), (f) period-3 (LLR), and (g) period-4 (LLLR). Panels (b), (c), and (e) illustrate $\ell\text{RPC}$ for fractional durations of UPOs, which also align with secondary RPC peaks, and are related to orbits trapped in a single wing.  
The recurrence plot was constructed with a threshold such that only \(1\%\) of the trajectory points are recurrent on average.
}
    \label{fig:lRPC_lorenz}
\end{figure}

To evaluate our quantifier in a three-dimensional chaotic system, we compute the global RPC for the Lorenz '63 system using a simple vertical line motif, defined as \(\textbf{w}_{\Delta i, \Delta j} = \delta_{\Delta i, 0}\delta_{\Delta j, \Delta t/dt}\), while varying the time lag \(\Delta t\). The time lag \(\Delta t\) is related to the sampling interval \(dt\) by \(\Delta t = k\,dt\), where \(k\) denotes the number of steps in the time series.
 The RPC as a function of the time lags presents distinctive peaks [Fig. ~\ref{fig:lRPC_lorenz}(a)]. 
 
The involution symmetry of the Lorenz '63 system gives rise to rich dynamics that generate intrinsic characteristic times. These intrinsic timescales span a broad frequency spectrum, which, in turn, impacts the dynamics, i.e., by making phase locking to external periodic forcing particularly challenging~\cite{park1999phase, zaks1999alternating}. The alternating wing dynamics is typically described symbolically by the system's itinerancy over the wings, using sequences such as \( \text{LLR} \), which reflect transitions between the attractor lobes~\cite{LI2024114751_UPO_Lorenz}. These symbolic sequences are also used to distinguish different UPOs; for instance, \( \text{LLR} \) corresponds to a period-three orbit.

The time lags related to correlation peaks correspond with precision to the well-documented periods of the shortest UPOs in the Lorenz system~\cite{LI2024114751_UPO_Lorenz}. This provides a quantitative validation of RPC as a powerful tool for UPO detection. For comparison, we also plot the standard autocorrelation function, which fails to identify these structurally significant periods as it measures the linear co-variation of the signal rather than the geometric recurrence of states in phase space.

We use $\ell\text{RPC}$ to visualize UPOs in phase space. For each significant time lag $\Delta t$ identified in Fig.~\ref{fig:lRPC_lorenz}(a), we compute the  $\ell\text{RPC}$ for every point along the attractor's trajectory. The results are shown in Figs.~\ref {fig:lRPC_lorenz}(b-f). In these plots, the Lorenz attractor is colored according to the local $\ell\text{RPC}$ value. The regions of highest correlation correspond to the geometric paths of the UPOs associated with the selected period $\Delta t$.
We observe two distinct classes of orbits. First, there are orbits confined entirely within a single wing of the attractor, associated with shorter-period homoclinic structures. Second, we identify orbits that traverse between the two lobes, corresponding to the well-known heteroclinic unstable periodic orbits, such as LR and LLR.

The $\ell\text{RPC}$ serves as an effective tool for visualizing the underlying structure of chaotic dynamics. Specifically, it reveals that states in the vicinity of a UPO exhibit recurrence patterns that are strongly correlated with a time lag corresponding to the orbit's period. Consequently, the $\ell\text{RPC}$ functions both as a detector and a visualization framework for uncovering the hidden skeleton of the chaotic attractor.

Correlated recurrences are shown here to be directly tied to the topological organization of chaos and to specific regions of phase space. This indicates that recurrent nonlinear dynamics are more effectively characterized through local properties rather than global statistics. Time-indexed recurrence pattern quantifiers therefore provide a complementary and sensitive perspective for revealing the fine-scale geometric structures embedded in recurrence plots.

Our results indicate that different dynamical regimes generate distinct recurrence motifs, implying that the weight matrix acts as a tunable lens for emphasizing system-specific geometry. Rather than selecting motifs manually, an important extension is to identify weight matrices that enhance contrast for a given system or highlight particular dynamical features. Preliminary evidence suggests that simple searches over candidate motifs already recover structures consistent with UPO periods or invariant-manifold geometry, motivating a data-driven strategy for inferring the intrinsic recurrence organization. Because $\ell\text{RPC}$ is highly sensitive to the spatial arrangement of recurrence points, it supports broader automated exploration of motif space without prescribing matrix forms in advance. Such inferred motifs would complement traditional line-based measures, extending the range of detectable structures arising from nonlinear evolution.

\section{\label{sec:conc}Conclusions}

We introduced Recurrence Pattern Correlation (RPC), a novel metric inspired by Moran's I spatial correlation measure, designed to quantify the correlation of recurrence patterns in time series data. Unlike traditional global metrics, RPC -- along with its localized, time-indexed variant ($\ell\text{RPC}$) -- provides a flexible framework that allows for the use of a user-defined correlation structure, enabling a detailed examination of rich, localized patterns within recurrence plots.

Our primary contribution is to demonstrate that recurrence patterns encode fundamental properties of the underlying dynamical system that extend beyond traditional recurrence-based quantifiers. The method's adaptive sensitivity, which emphasizes structure in sparse RPs and non-recurrence in dense RPs, allows for a robust analysis across different dynamical regimes

We validated the utility of RPC on a range of representative systems.
For stochastic processes like Gaussian white noise and Autoregressive, RPC correctly identified low intrinsic structure and patterns consistent with linear memory decay.
For the Logistic Map,  $\ell\text{RPC}$ effectively visualized the bifurcation diagram, with correlation patterns delineating the periodic windows and chaotic regions.
For the Standard Map,  $\ell\text{RPC}$ served as a powerful lens to dissect the mixed phase space, selectively highlighting the chaotic sea or the invariant structures of stable islands depending on the chosen correlation motif.
Most significantly, for Lorenz'63 system, we established a direct link between long-range correlations in recurrence patterns and the attractor's skeleton of Unstable Periodic Orbits (UPOs).  $\ell\text{RPC}$ not only identified the periods of the shortest UPOs but also precisely visualized their geometric paths in phase space.

Future work should focus on developing agnostic strategies to infer weight matrices that represent the system’s preferred recurrence motifs, rather than selecting them a priori. Combined with multivariate and cross-recurrence formulations, this would extend RPC toward detecting structured, time-lagged interactions across variables, enabling a compact geometric characterization of directional influence in complex nonlinear systems.

By generalizing recurrence pattern analysis and connecting it to foundational concepts in chaos theory, this work introduces Recurrence Pattern Correlation to reveal the structural organization of a system's recurrences. The local version, $\ell\text{RPC}$, is particularly noteworthy as it enhances interpretability by directly linking recurrence patterns to locations in the state space. This provides a powerful and flexible tool for simultaneously extracting both temporal and geometric information from recurring dynamics.

\section*{Code Availability} 
The Julia code developed for the Recurrence Pattern Correlation analysis is openly available in the GitHub repository, \textit{RecurrencePatternsCorrelation}, at \url{https://github.com/GabrielMarghoti/RecurrencePatternsCorrelation.git}. The specific version of the code used to produce the results in this paper (v1.0) is permanently archived on Zenodo at \url{https://doi.org/10.5281/zenodo.15855176}.

\section*{Acknowledgments}
This study was financed in part by the Coordena\c{c}\~{a}o de Aper\-fei\c{c}oamento de Pessoal de N\'{i}vel Superior (CAPES), Brazil, Finance Code 001, through project No. 88887\-.989388\-/2024\--00, and the PROBRAL project No. 88881\-.895032\-/2023\--01; by the Conselho Nacional de Desenvolvimento Cient\'ifico e Tecnológico (CNPq), Brazil, under Grants Nos. 305189\-/2022\--0, 407072/2022-5, 408254\-/2022\--0, and 300064\-/2023\--3; and by the Funda\c{c}\~{a}o de Amparo à Pesquisa do Estado de S\~{a}o Paulo (FAPESP), Brazil, under Grants Nos. 2023\-/07704\--5 and 2024\-/22136\--6.

\section*{Author Contribution Statement}

\textbf{G. M.:} Conceptualization, Investigation, Visualization, Methodology, Writing - original draft.
\textbf{M. P. S.:} Investigation, Visualization,  Writing - review \& editing.
\textbf{T. L. P.:} Supervision, Writing - review \& editing.
\textbf{S. R. L.:} Supervision, Writing - review \& editing.
\textbf{J. K.:} Supervision, Writing - review \& editing.
\textbf{N. M.:} Conceptualization, Supervision, Writing - review \& editing.

\section*{Declaration of Competing Interest}

The authors declare that they have no conflicts of interest.

\bibliographystyle{model1-num-names}
\bibliography{references.bib}

\end{document}